\documentclass[a4paper,11pt]{article}

\usepackage{jheppub} 
\usepackage{amsmath,amsfonts,amssymb}
\usepackage{mathtools}
\usepackage{ascmac}
\usepackage{physics}
\usepackage{color}

\def\bs#1{\boldsymbol{ #1 }} 
\newcommand{\wt}{\widetilde}
\newcommand{\wh}{\widehat}
\newcommand{\ol}{\overline}
\newcommand{\del}{\partial}
\newcommand{\ra}{\rightarrow}

\newcommand{\nn}{\nonumber}

\numberwithin{equation}{section}

\newcommand{\nV}{n_\mathrm{V}}

\title{\boldmath Chiral magnetic wave and axial charge relaxation in flavor branes}
\author[a,b]{Masataka Matsumoto}
\author[a]{and Kensei Tanaka}%
\affiliation[a]{%
 Department of Physics, Chuo University,\\ 1-13-27 Kasuga, Bunkyo-ku, Tokyo 112-8551, Japan}
\affiliation[b]{Department of Physics, Shanghai University,\\ 99 Shangda Road, Shanghai 200444, China}
\emailAdd{mmatsumoto173@g.chuo-u.ac.jp}
\emailAdd{a20.p3h8@g.chuo-u.ac.jp}

\abstract{  
We investigate the chiral magnetic wave and axial charge relaxation in the chiral symmetry broken phase of the holographic D3/D7 model with massless quarks, where the axial charge in the flavor sector can relax through transfer to the adjoint sector in the probe limit. We find that the dispersion relations of the quasi-normal modes are well described by the phenomenological expression for the chiral magnetic wave when the relaxation rate is small. We then determine how this rate depends on vector charge density and the chiral condensate. At higher temperatures, it increases with charge density and decreases with increasing condensate, while these dependences become non-monotonic at lower temperatures. These results show that the phenomenological description of the chiral magnetic wave also applies in the chiral symmetry broken phase in the quasi-hydrodynamic regime.
}

\begin{document}
\maketitle
\flushbottom


\section{Introduction}
Fermions can be classified into left- and right-handed particles according to their chirality, which coincides with helicity, in the massless limit.
In ordinary circumstances, the populations of left- and right-handed fermions are typically balanced.
However, a chiral imbalance can arise under extreme situations, such as quark-gluon plasmas and nuclear matters generated in heavy-ion collisions \cite{Kamada:2022nyt}.
In such a relativistic chiral matter, anomalous transport phenomena associated with the chiral anomaly are expected to be observed.

The chiral imbalance can be relaxed through chirality-flipping interactions between left- and right-handed fermions via the fermion mass term. 
In the massless limit, the system possesses chiral symmetry, which is explicitly broken by a finite fermion mass.
Chiral symmetry can also be spontaneously broken, while the chiral anomaly leads to the nonconservation of the axial current in the presence of the electric and magnetic fields \cite{PhysRev.177.2426,Bell:348417}, given by
\begin{equation}\label{eq:anomaly}
    \del _\mu j^\mu _\mathrm{A}=\frac{e^{2}}{2\pi^2} \vec{E}\cdot\vec{B},
\end{equation}
where $j_\mathrm{A}^\mu=(n_\mathrm{A},\vec{j}_\mathrm{A})$ denotes the four-vector of the axial current.
Assuming spatially homogeneous systems ($\vec{\nabla} \cdot \vec{j}_{\rm A} =0$), Eq.~\eqref{eq:anomaly} is interpreted as the generation of the axial charge by the electromagnetic fields.
Additionally, in the presence of the scattering between left- and right-handed fermions, the axial charge can be relaxed even if the fermions are massless, which is phenomenologically described as
\begin{equation}\label{eq:AnomaryDisspation}
    \del _t n _\mathrm{A}=\frac{e^{2}}{2\pi^2} \vec{E}\cdot\vec{B}-\frac{n_\mathrm{A}}{\tau},
\end{equation}
where $\tau$ denotes the relaxation time of the axial charge and depends on the details of the system.

Typical examples of the chiral transport phenomena are chiral magnetic effect \cite{PhysRevD.22.3080,Fukushima:2008xe} and chiral separation effect \cite{Metlitski:2005pr,Son:2009tf}, in which vector and axial currents are induced by the external magnetic field in the presence of the corresponding chemical potentials.
As proposed in \cite{Kharzeev:2010gd}, the interplay of these anomalous transports leads to the coupling between the vector and axial current fluctuations, giving rise to the collective excitations that exhibit a long-lived propagating wave, which is referred to as the {\it chiral magnetic wave} (CMW). 
The CMW has been studied within the framework of chiral kinetic theory, where the effects of chirality-flipping processes on the collective excitations have been investigated \cite{Stephanov:2015,Hanai:2022}.
The CMW has also been investigated experimentally in relativistic heavy-ion collisions \cite{Ke:2012,Belmont:2014,Adamczyk:2015}. Although some experimental results have suggested possible signatures of the CMW, its existence remains inconclusive due to background effects \cite{STAR:2023,ALICE:2023}.

In \cite{Kharzeev:2010gd}, the CMW at a strongly correlated system was explicitly demonstrated in the Sakai-Sugimoto model \cite{Sakai:2004cn,Sakai:2005yt} using the gauge/gravity duality.
The gauge/gravity duality (or holography) claims the equivalence between a classical gravity theory and a strongly-correlated gauge theory \cite{Maldacena:1997re,Gubser:1998bc,Witten:1998qj}.
The holographic approach is a powerful tool to analyze quantum many-body systems with a strong coupling via the classical analysis of a gravitational system, including various applications to condensed matter physics \cite{zaanen2015holographic,ammon2015gauge,hartnoll2018holographic}, QCD \cite{kim2013holographic,casalderrey2014gauge}, nonequilibrium phenomena \cite{hubeny2010holographic,liu2020holographic}, and so on.
The aforementioned Sakai-Sugimoto model is a top-down holographic model which consists of the D-branes, specifically D4- and D8-branes, and represents the strongly correlated QCD system \cite{Sakai:2004cn,Sakai:2005yt}. 
Another well-known top-down model is the D3/D7 model in which the chiral properties, such as the chiral transport phenomena \cite{Hoyos:2011us} and the spontaneous chiral symmetry breaking \cite{Babington:2003vm,Filev:2009xp,Evans:2010iy}, have been extensively studied.
Within bottom-up holographic models, the chiral magnetic wave and related anomalous transport phenomena have been investigated, with particular attention to axial charge relaxation and the dynamics of chiral magnetic currents \cite{Jimenez-Alba:2014iia,Grieninger:2023myf,Grieninger:2023wuq,Ahn:2024ozz}.

In this paper, we investigate the CMW in the D3/D7 model.
In contrast to the previous study based on the Sakai-Sugimoto model in the chirally symmetric phase \cite{Kharzeev:2010gd}, we focus on collective excitations in the chiral symmetry broken phase. 
In this phase, the spontaneous breaking of axial symmetry gives rise to a Nambu-Goldstone (NG) mode, which contributes to axial current fluctuations and can modify the low-energy collective dynamics through its coupling to vector current fluctuations. 
As shown in Ref.~\cite{Ishigaki:2020vtr}, the NG mode exhibits dissipative dynamics. 
It is therefore nontrivial whether the resulting collective excitations can still be described by the phenomenological dispersion relation of the CMW.

To address this question, we investigate the dispersion relations of the collective modes, with particular focus on the axial charge relaxation time $\tau$.
In particular, we study the dependence on the background charge density and the chiral condensate, corresponding to the strength of the chiral symmetry breaking, and the consistency with the hydrodynamic predictions.
We note that the effects of finite quark mass on axial charge dynamics and chiral separation effect have previously been studied in the D3/D7 model \cite{Guo:2016nnq,Guo:2016dnm}, whereas we restrict our analysis to the massless quarks.

This paper is organized as follows.
In section \ref{sec:2}, we briefly review the CMW and a phenomenological description of its dispersion relation in the presence of the dissipation of the axial charge.
In section \ref{sec:3}, we explain our holographic setup.
We introduce the chiral symmetry breaking background in the presence of the external magnetic field and charge density at finite temperature in the D3/D7 model.
Section \ref{sec:4} is the main part of this paper, where we consider the fluctuations and numerically compute the quasi-normal modes that correspond to the dispersion relation of the chiral magnetic wave.
We also investigate the dependence of the relaxation rate as a function of the background charge density and the chiral condensate.
Section \ref{sec:5} is devoted to conclusion and discussion.


\section{Review: Chiral Magnetic Wave}\label{sec:2}
Chiral transport phenomena, such as chiral magnetic effect (CME) and chiral separation effect (CSE), are quantum transport processes caused by the chirality of carrier particles.
CME induces an electric (vector) current along an external magnetic field $\bs{B}$ mediated by a finite axial charge density $n_{\rm A}$, corresponding to a chiral imbalance, in the absence of an external electric field, given by \cite{PhysRevD.22.3080, Fukushima:2008xe}
\begin{equation}
  \vec{j}_\mathrm{V}=\frac{\mu_\mathrm{A}}{2 \pi ^2}\vec{B}, \label{eq:CME}
\end{equation}
where $\mu_{\rm A}$ is the axial chemical potential.
In contrast, CSE induces an axial current in the presence of a finite charge density $n_{\rm V}$ as \cite{Metlitski:2005pr,Son:2009tf}
\begin{equation}
  \vec{j}_\mathrm{A}=\frac{\mu_\mathrm{V}}{2 \pi ^2}\vec{B} \label{eq:CSE},
\end{equation}
where $\mu_{\rm V}$ is the chemical potential.
In general, their spatial distribution induces each diffusive current, i.e. Fick's law, with a diffusion constant $D$\footnote{Strictly speaking, the diffusion constant should be distinguished in the vector and axial sector as $D_{\rm V}$ and $D_{\rm A}$. We here assume they are identical $D_{\rm V}=D_{\rm A}$ for simplicity.}. 
Then one can write each current as
\begin{align}
  \vec{j}_{\mathrm{V}}=\frac{\vec{B}}{2 \pi ^2 \chi_{\rm A}}n_\mathrm{A} - D \,\vec{\nabla} n_\mathrm{V}, \label{eq:cme}\\
  \vec{j}_{\mathrm{A}}=\frac{\vec{B}}{2 \pi ^2 \chi_{\rm V}}n_\mathrm{V}- D \,\vec{\nabla}  n_\mathrm{A}. \label{eq:cse}
\end{align}
Here, $n_{\rm V, A}$ and $\mu_{\rm V,A}$ denote fluctuations about the background values, and we assume the linear relation, i.e.
\begin{equation}
    \begin{pmatrix}
    \mu_{\rm V} \\ \mu_{\rm A}
    \end{pmatrix} =
    \begin{pmatrix}
    \chi_{\rm V}^{-1} & 0 \\ 0 & \chi_{\rm A}^{-1} 
    \end{pmatrix}
    \begin{pmatrix}
    n_{\rm V} \\ n_{\rm A}
    \end{pmatrix},
\end{equation}
with the charge susceptibility $\chi_{\rm V}$ and the axial charge susceptibility $\chi_{\rm A}$.
Note that the off-diagonal components vanish by imposing parity invariance.
These susceptibilities become identical if the system is in the chiral symmetric phase \cite{Kharzeev:2010gd}.

The vector and axial four-currents are conserved if the system possesses the corresponding $U(1)_{\rm V}$ and $U(1)_{\rm A}$ symmetries, respectively.
However, axial symmetry can be broken explicitly by fermion mass, spontaneously by vacuum, and by anomaly.
Explicit breaking and the axial anomaly lead to the non-conservation of the charge density.
Writing the relaxation time as $\tau$, the divergence of the vector and axial currents can be expressed as
\begin{eqnarray}
  \del_\mu {j^{\mu}_\mathrm{V}} &=&\del_t n_\mathrm{V} + \vec{\nabla}\cdot \vec{j}_{\mathrm{V}}=0, \label{eq:cons}\\
  \del_\mu {j^{\mu}_\mathrm{A}} &=& \del_t n_\mathrm{A} + \vec{\nabla}\cdot \vec{j}_{\mathrm{A}}=-\frac{1}{\tau} n_\mathrm{A}. \label{eq:ncons}
\end{eqnarray}
Note that this form of the non-conservation of the axial charge density is a phenomenological expression, which may arise from weak $U(1)_{\rm A}$ symmetry breaking or charge transfer to the other degrees of freedom.
Combining them with the constitutive relations \eqref{eq:cme} and \eqref{eq:cse}, we obtain the coupled equations for $n_{\rm V}$ and $n_{\rm A}$ as
\begin{equation}
    \begin{pmatrix}
        \del_{t} - D\nabla^{2} & \quad \kappa \chi_{\rm A}^{-1}\vec{B}\cdot\vec{\nabla} \\
        \kappa \chi_{\rm V}^{-1}\vec{B}\cdot\vec{\nabla} & \quad \del_{t} +\tau^{-1} - D\nabla^{2}
    \end{pmatrix}
    \begin{pmatrix}
        n_{\rm V} \\ n_{\rm A} 
    \end{pmatrix}=0,
\end{equation}
where $\kappa =1/(2\pi^{2})$.
In Fourier space, we have
\begin{equation}
    \begin{pmatrix}
        -i\omega + Dk^{2} & \quad i \kappa \chi_{\rm A}^{-1}\vec{B}\cdot\vec{k} \\
        i\kappa \chi_{\rm V}^{-1}\vec{B}\cdot\vec{k} & \quad -i\omega+\tau^{-1} + Dk^{2}
    \end{pmatrix}
    \begin{pmatrix}
        \tilde{n}_{\rm V} \\ \tilde{n}_{\rm A} 
    \end{pmatrix}=0,
\end{equation}
where $k^{2} = |\vec{k}|^{2}$ and $\tilde{n}_{\rm V/A}$ are the corresponding Fourier modes.
Thus, the dispersion relation for the collective modes is given by
\begin{equation}\label{eq:CMW0}
  \omega_{\pm}= - \frac{i}{2\tau} -i D k^2  \pm \sqrt{\frac{\kappa^{2}\left(\vec{B}\cdot \vec{k} \right)^2}{\chi_{\rm V}\chi_{\rm A}}-\frac{1}{4\tau^2}}.
\end{equation}
Considering an external magnetic field along the $z$ direction, $\vec{B} = (0,0,B)$, and focusing on the longitudinal dispersion relation of $k_z$, we identify a characteristic momentum scale:
\begin{equation}
    k_{c} = \frac{\sqrt{\chi_{\rm V} \chi_{\rm A}}}{2\kappa\tau B },
\end{equation}
above which the fluctuations exhibit damped oscillatory behavior and below which they are overdamped.
In the small-$k_z$ limit, one mode corresponds to the massless diffusion mode $\omega_{+} \simeq - i \wt{D} k_z^2$ with $\wt{D} = D +(4\tau k_{c}^{2})^{-1}$, and the other corresponds to the gapped mode, $\omega_{-} \simeq -i/\tau$.
In the absence of the axial charge relaxation, i.e. $\tau \to \infty$, the dispersion relation reduces to
\begin{equation}
    \omega_{\rm CMW} =  \pm v k_{z} - i D k_{z}^{2}, \qquad v = \frac{\kappa B}{\sqrt{\chi_{\rm V}\chi_{\rm A}}},
\end{equation}
which corresponds to the chiral magnetic wave (CMW) \cite{Kharzeev:2010gd,Burnier:2011bf}, a propagating mode subject to diffusive damping.

In this paper, we demonstrate that the collective modes of interest in our holographic model obey the dispersion relation of \eqref{eq:CMW0} in the hydrodynamic regime.
As we will discuss in the later section, the parameters in the dispersion relation, such as $\tau$ and $D$, depend on the background parameters, such as temperature and charge density.


\section{Holographic Setup: D3/D7 Model}\label{sec:3}
In this section, we briefly review our holographic setup based on the D3/D7 model.
We consider the system at finite temperature and charge density in the presence of an external magnetic field.
For massless quarks, this model is known to exhibit spontaneous chiral symmetry breaking in certain parameter regions \cite{Evans:2010iy}.
\subsection{Setup}\label{subsec:31}
To analyze strongly coupled fermionic systems, we consider the D3/D7 model \cite{Karch:2002sh}.
The D3/D7 model is a top-down holographic model consisting of $N_c $ D3-branes and $N_f$ D7-branes. In this paper, we set the number of flavors to one, i.e. $N_f =1$.
Table~\ref{tab:D3D7} summarizes the directions along which the D-branes extend.

\begin{table}[h]
  \centering
  \caption{ Intersection of the D3-branes and D7-brane in (9+1)-dimensional spacetime.
  }\label{tab:D3D7}
  \begin{tabular}{|c||c|c|c|c|c|c|c|c|c|c|}
      \hline
       & $X^0$ & $X^1$  & $X^2$ & $X^3$ & $X^4$ & $X^5$ & $X^6$ & $X^7$ & $X^8$& $X^9$\\ \hline
      D3 & \checkmark  & \checkmark  & \checkmark  & \checkmark  &  & &  &  & &  \\ \hline
      D7 & \checkmark  & \checkmark  & \checkmark  & \checkmark  & \checkmark  & \checkmark  & \checkmark & \checkmark  &  &   \\ \hline
  \end{tabular}
\end{table}
We first summarize the field theory realized in the D3/D7-brane system in $(9+1)$-dimensional flat spacetime.
On the D3-brane worldvolume, open strings with both endpoints attached to the D3-branes yield the $U(N_c)$ $\mathcal{N}=4$ super Yang-Mills theory ($\mathcal{N}=4$ SYM).
The Yang-Mills coupling constant is given by $g_\mathrm{YM}^2 = 2 \pi g_\mathrm{s}$, where $g_\mathrm{s}$ is the closed string coupling constant. 
The effective coupling constant of the gauge theory is given by the 't~Hooft coupling $\lambda= g_\mathrm{YM}^2 N_c$. 
The $\mathcal{N}=4$ SYM theory contains a gauge field, four Weyl fermions, and six real scalar fields, which are transformed in the adjoint representation of $U(N_{c})$.
This theory has an $SO(6)$ R-symmetry that rotates these six real scalar fields.
To impose a flavor sector, we consider open strings with their endpoints attached to the D3-branes and D7-brane.
These open strings give the $\mathcal{N}=2$ hypermultiplet containing a Dirac fermion and two complex scalars.

Next, we discuss symmetries in the theory.
Due to the D-brane intersection, the $SO(6)$ R-symmetry of $\mathcal{N}=4$ SYM is broken to $SO(4)\times U(1)$.
Here $U(1) \simeq SO(2)$ corresponds to a rotational symmetry in the plane of $(X^8,X^9)$ coordinates.
This symmetry is explicitly broken if the D7-brane is separated from the D3-branes, and the distance between them corresponds to the mass of the flavor fermions.
Since the corresponding transformation acts as a $U(1)$ axial transformation for the Dirac fermions in the $\mathcal{N}=2$ hypermultiplet, the separation of the D3-branes and D7-brane corresponds to the explicit breaking of the $U(1)$ axial symmetry in the dual field theory. Hereafter, we denote this axial symmetry as $U(1)_{\rm A}$. 
Note that the $U(1)_{\rm A}$ axial charge considered here is carried by the flavor sector and can relax through transfer to the adjoint sector \cite{Hoyos:2011us,Guo:2016nnq}, which acts as a reservoir in the probe limit. Thus, the axial charge in the flavor sector need not be conserved even for massless quarks.

Let us now consider the dual gravitational background of the D-brane system.
Taking the large $N_c$ limit $N_c\gg 1 $ with $\lambda$ fixed, and the large 't~Hooft limit $\lambda = g_\mathrm{YM}^2 N_c\gg1 $, the gravitational dual of the D3/D7 model is the classical type IIB supergravity with a probe D7-brane, where we also assume the probe limit.
The gravitational background of D3-brane is given by the AdS$_5$-Schwarzschild black hole times $S^5$, whose metric is given by
\begin{gather}\label{eq:backgroundMet}
  \dd s^2 = \frac{1}{u^2}\left( - f(u) \dd t^2 + \dd \vec{x}^2 + \frac{\dd u^2}{f(u)} \right) + \dd \theta^2  + \sin ^2\theta \dd \phi^{2} + \cos ^2 \theta \dd \Omega_3^2, \\
  f(u)= 1- \frac{u^4}{u_\mathrm{H}^4},
\end{gather}
where we have set the AdS radius to unity.
Here, $(t,\vec{x})$ denote the $(3+1)$-dimensional coordinates, $u$ is the AdS radial coordinate, and $(\theta,\phi,\Omega_3)$ parametrize the $S^5$.
The AdS boundary is located at $u \to 0$, and $u = u_H$ corresponds to the black hole horizon.
Here, $\dd \Omega_3$ is the line element of $S^3$, which is wrapped by the D7-branes.
The Hawking temperature is given by $T=1/(\pi u_\mathrm{H})$.
In addition, the background contains a Ramond-Ramond (RR) 4-form potential:
\begin{equation}
  C_4= \frac{1}{u^4}\dd t  \wedge \dd x \wedge \dd y \wedge \dd z + \cos ^4 \theta \dd \phi \wedge \omega(S^3),
\end{equation}
where $\omega(S^3)$ is the volume element of $S^3$.\footnote{There are other fields such as the dilaton and the Kalb-Ramond field. However the dilaton is constant in this case, and the Kalb-Ramond field is completely decoupled from the gauge theory we considered.}

The D7-brane dynamics is determined by the sum of the Dirac-Born-Infeld (DBI) action and the Wess-Zumino term:
\begin{align}
  S_\mathrm{D7}&=S_\mathrm{DBI}+S_\mathrm{WZ}, \label{eq:D7action}\\
  S_\mathrm{DBI}&=- T_\mathrm{D7} \int \dd ^8 \xi \sqrt{-\det (g_{ab}+ 2\pi \alpha' F_{ab})}, \label{eq:DBIaction}\\
  S_\mathrm{WZ}&=\frac{(2\pi \alpha')^2}{2} T_\mathrm{D7} \int P[C_4]\wedge F \wedge F, \label{eq:WZaction}
\end{align}
where $\alpha'=l_{\rm s}^{2}$ with the string length $l_{\rm s}$ and the $T_\mathrm{D7}$ is the tension of D7-brane given by $T_\mathrm{D7}=(2\pi )^{-7}g_\mathrm{s}^{-1}\alpha^{\prime -4}$.
$g_{ab}$ is the induced metric of D7-brane defined as
\begin{equation}
  g_{ab} = \frac{\del X^M}{\del \xi^a}\frac{\del X^N}{\del \xi^b} G_{MN}
\end{equation}
where $X^M\, (M=0,1,\cdots 9)$ are the coordinates of target space, $G_{MN}$ is the background metric (\ref{eq:backgroundMet}) and  $\xi^a(a=0,1,\cdots 7)$ are the D7-brane world volume coordinates.
$F_{ab} = \partial_{a}A_{b} - \partial_{b}A_{a}$ is the field strength of $U(1)_\mathrm{V}$ gauge field $A_{a}$ and $P[C_4]$ is the pullback of the Ramond-Ramond field to D7-brane worldvolume.
In our work we employ the static gauge and we set $2\pi \alpha'=1$ for simplicity.


\subsection{Background Solutions} \label{subsec:32}
To compute CMW in this setup, we need to construct a background solution upon which fluctuations are imposed. For our purpose, we consider the following ansatz:
\begin{equation}\label{eq:ansatz1}
  \theta = \theta(u) ,\quad \phi =0,\quad A_t= a_t (u),\quad A_y = B x,\quad A_x=A_z=0,
\end{equation}
where we refer to the scalar field $\theta (u)$ as the embedding function. 
Since the D7-brane action (\ref{eq:D7action}) depends on $a_t'(u)$ but not on $a_t(u)$, the quantity $\partial \mathcal{L}_\mathrm{D7}/\partial a_t'(u)$ is conserved along the AdS direction.
According to the holographic dictionary \cite{Karch:2007pd}, the negative of this conserved quantity corresponds to the $U(1)_\mathrm{V}$ charge density $\nV$.

Performing the Legendre transformation with respect to $a_t(u)$, we obtain the following action:
\begin{align}\label{eq:eomoftheta}
  \wt{S}_\mathrm{D7}&= S_\mathrm{D7} - \int \dd ^8 \xi \,a_t'(u)\frac{\delta S_\mathrm{D7}}{\delta a_t'(u)}\nn \\
  &=-\int \dd^{4}x \,\dd u \, \frac{\sqrt{\left(n_\mathrm{V}^2 u^6 + (1 + B^2 u^4) \mathcal{N}^2 \cos^6\theta(u)\right) \left(u_\mathrm{H}^4 + u^2 (-u^4 + u_\mathrm{H}^4) \theta'(u)^2\right)}}{u^5 u_\mathrm{H}^2},
\end{align}
where $n_\mathrm{V}=-\del \mathcal{L}_\mathrm{D7}/\del a_t'(u)$ and $\mathcal{N}= T_\mathrm{D7} (2\pi^2)= N_c/(2\pi)^2$.
The equation of motion for $\theta(u)$ derived from (\ref{eq:eomoftheta}) is a highly non-linear differential equation, which can be solved numerically to obtain the background embedding function.
If we consider the asymptotic behavior of $\theta (u)$ near the AdS boundary, it is expanded as
\begin{equation}
  \theta(u)= m_q u + \theta_2 u^3 +\cdots. \label{eq:theta_asymp}
\end{equation}
According to the holographic dictionary \cite{Karch:2007pd}, the coefficient $m_q$ corresponds to the flavor mass and the chiral condensate is given by
\begin{equation}\label{eq:CC}
  \langle \ol{\psi }\psi\rangle = \mathcal{N}\left(2\theta_2 - \frac{m_q^3}{3}\right).
\end{equation}

We solve the equation of motion by the shooting method and obtain the numerical solutions of the embedding function $\theta(u)$.
In this paper, we set $\mathcal{N}=1$ in addition to $2\pi\alpha'=1$, which correspond to $\lambda=2\pi g_{s}N_{c}=2\pi^2$ and $N_c=(2\pi)^2$ in the dual field theory, where we have used the relation $4\pi g_sN_c\alpha'^2=1$.
The background solution $\theta(u)$ is then characterized by three parameters: the magnetic field $B$, the charge density $n_\mathrm{V}$, and the temperature $T$ (or horizon radius $u_\mathrm{H}$).
By using the scaling invariance of the theory, we choose to normalize all physical quantities by the magnetic field $B$.
Consequently, the solution $\theta(u)$ is fully specified by two dimensionless parameters: $(\pi T/\sqrt{2 B}, n _\mathrm{V}/B^{3/2})$.

\begin{figure}[thbp]
\begin{center}
\includegraphics[width=.85\textwidth]{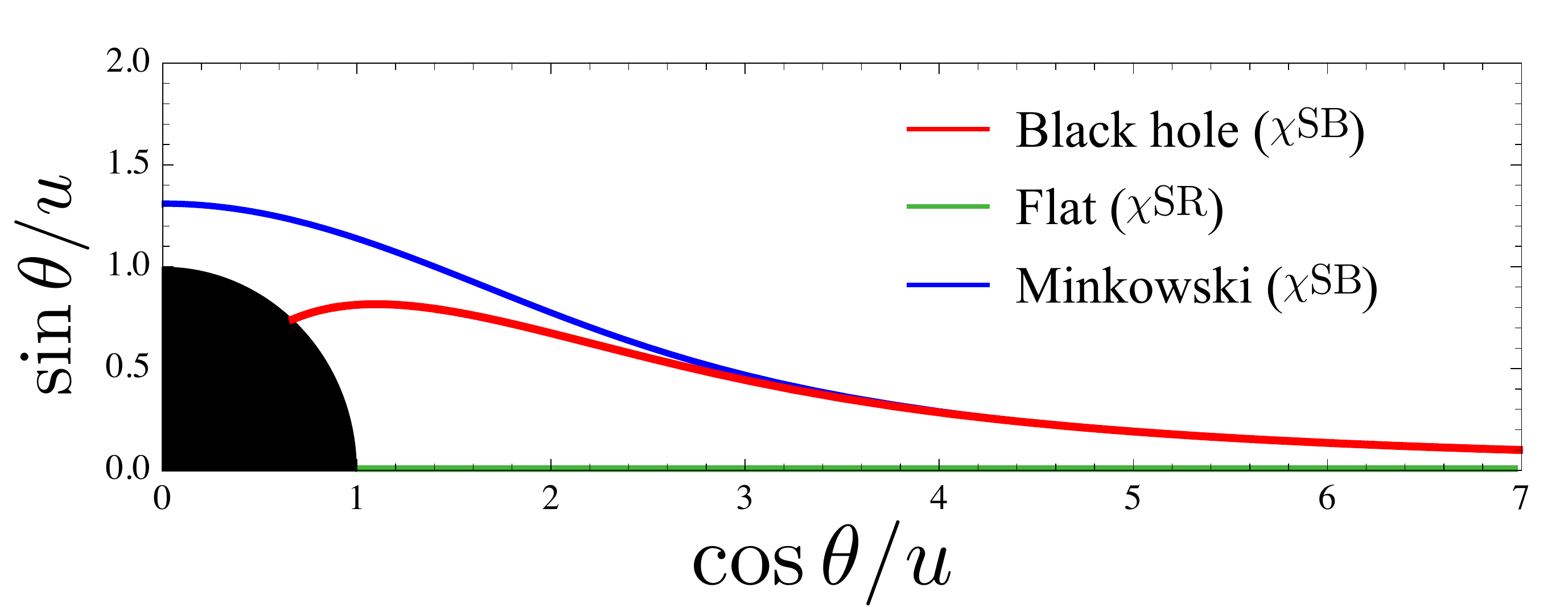}
\caption{The typical configurations of the D7-brane embedding function $\theta(u)$ for different types of the embedding.
The black quarter circle denotes the black hole with $u_{\rm H} =1$.
The AdS boundary locates at the infinity of the horizontal axis, and we consider the configurations which asymptotically approach the horizontal axis, corresponding to $m_q = 0$ from \eqref{eq:theta_asymp}. In the black hole embedding, we here refer to the trivial solution $\theta(u) =0$ as ``Flat" (green) and the non-trivial solution $\theta(u)\neq 0$ as ``Black hole" (red), corresponding to the $\chi$SR and $\chi$SB phases, respectively. The Minkowski embedding solution is denoted by the blue curve.}
\label{fig:embedding}
\end{center}
\end{figure}

To solve the equation of motion for $\theta (u)$, we need to impose two boundary conditions.
Since we focus on the massless solutions, we need to impose the boundary condition such that $m_q=0$ at the AdS boundary.
Another boundary condition must be classified based on the type of solution.
There are two different possible solutions for the embedding function $\theta (u)$: the Minkowski embedding and the Black hole embedding.
In the black hole embedding, the D7-brane reaches the black hole horizon.
To avoid a singularity at the black hole horizon, the regularity condition is imposed on $\theta'(u_\mathrm{H})$, which is the other boundary condition to obtain the black hole embedding.
Among the black hole embeddings, the trivial solutions $\theta(u)=0$ represent chiral symmetry restored phase ($\chi$SR), whereas the non-trivial solutions $\theta(u) \neq 0$ exhibit a spontaneous chiral symmetry breaking ($\chi\text{SB}$) as the chiral symmetry is broken despite having massless fermions ($m_q=0$).
In the Minkowski embedding, the D7-brane does not reach the black hole horizon.
For this solution, we need to impose the boundary condition at $u=u_\mathrm{max}<u_{\rm H}$ 
to avoid conical singularity.
Figure~\ref{fig:embedding} shows the plots of the numerical solutions $\theta(u)$ for each embedding.

\begin{figure}[htbp]
\begin{center}
\includegraphics[width=.8\textwidth]{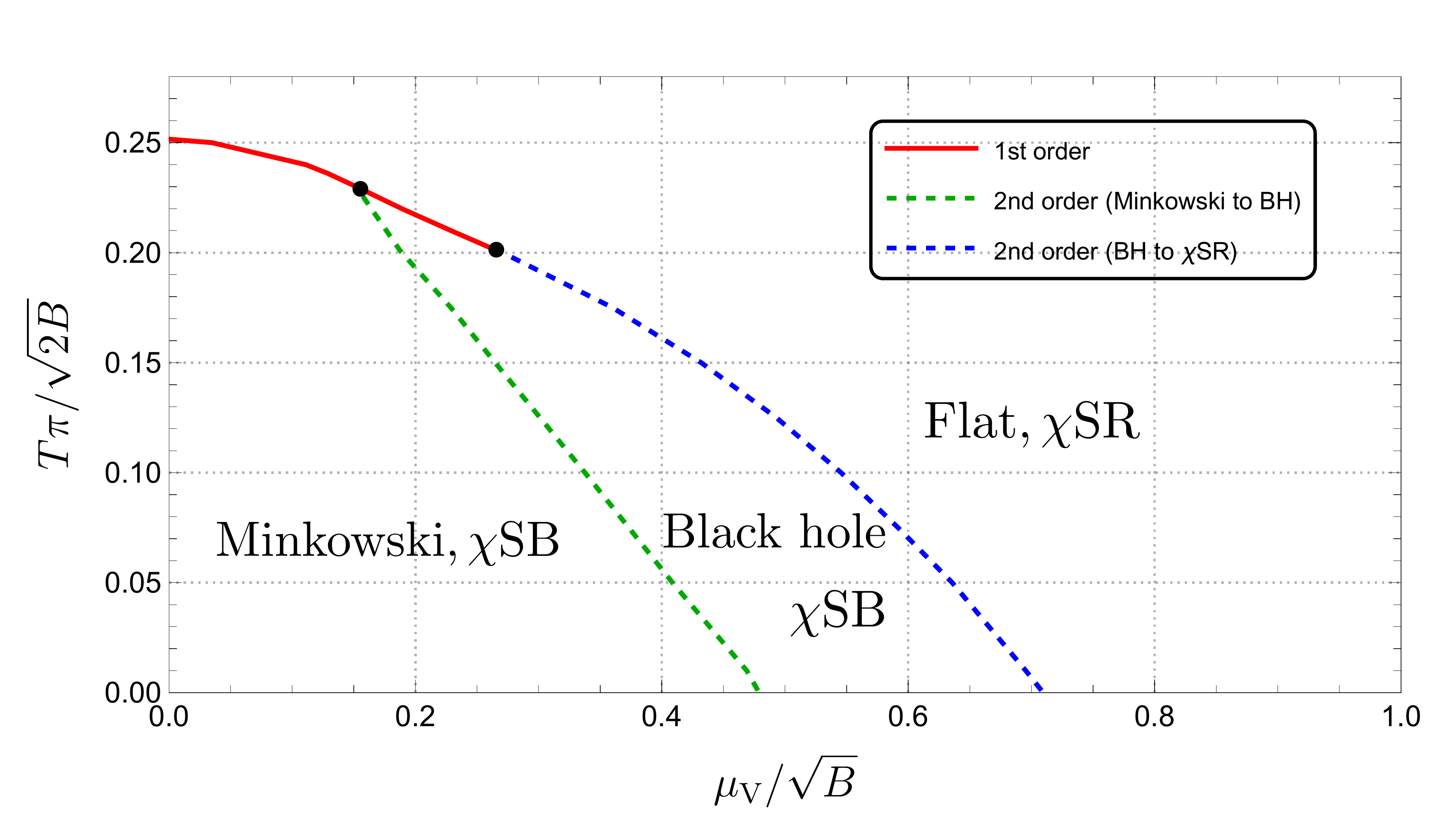}
\caption{Phase diagram for the massless solutions in the plane of $(\pi T/\sqrt{2 B}, \mu_\mathrm{V}/\sqrt{B})$. Each phase is separated by the first- (solid) or second-order (dashed) phase transition lines. The phase diagram is originally drawn in \cite{Evans:2010iy}.}
\label{fig:phasediagram}
\end{center}
\end{figure}

The phase transitions among these solutions are determined by comparing the thermodynamic potentials \cite{Evans:2010iy}, defined from the on-shell action as\footnote{For the Minkowski embedding, the interval of integration must be $[0,u_{\rm max}]$.}
\begin{equation}
  \Omega (T,\mu_{\rm V} )= \frac{-{S}_\mathrm{D7}}{\mathrm{Vol}_{\mathbb{R}^{1,3}}}= \int ^{u_\mathrm{H}}_{0}\dd u \,  \frac{(1 + B^2 u^4)  \cos^6\theta(u)}{u^5 u_\mathrm{H}^2} \sqrt{\frac{u_\mathrm{H}^4 + u^2 (-u^4 + u_\mathrm{H}^4) \theta'(u)^2}{n_\mathrm{V}^2 u^6 + (1 + B^2 u^4)  \cos^6\theta(u)}},
\end{equation}
where $\mathrm{Vol}_{\mathbb{R}^{1,3}}$ denotes the volume integral in $(3+1)$-dimensional spacetime.\footnote{Since the integrand diverges in the UV region ($u \to 0$), we must introduce a small UV cutoff $\epsilon_\mathrm{UV}$ and add the following counterterm to regularize the divergence \cite{Karch:2007pd},
\begin{align}
  L_\text{counter term}=\frac{1}{4}\frac{\mathcal{N}}{\epsilon_\mathrm{UV}^4}-\frac{1}{2}\frac{\mathcal{N}}{\epsilon_\mathrm{UV}^4}\theta(\epsilon_\mathrm{UV})^2+\frac{5}{12}\frac{\mathcal{N}}{\epsilon_\mathrm{UV}^4}\theta(\epsilon_\mathrm{UV})^4 -\frac{1}{2}\mathcal{N}(2\pi \alpha')^2 B^2 \log \epsilon_\mathrm{UV} \nn.
\end{align}
}
Since the normalized thermodynamic potential $\Omega/B^{2}$ is given as a function of the normalized parameters of $(\pi T/\sqrt{2B}, \mu_{\rm V}/\sqrt{B})$, we also compute the $U(1)_{\rm V}$ chemical potential for the black hole embedding by \cite{Kobayashi:2006sb,Mateos:2007vc,Evans:2010iy}
\begin{equation}
  \mu _\mathrm{V}= \lim_{u\to 0} a_t(u) = \int_{0}^{u_\mathrm{H}}\dd u\, \frac{n_\mathrm{V}\, u}{u_\mathrm{H}^2} \sqrt{\frac{u_\mathrm{H}^4 + (-u^6 + u^2 u_\mathrm{H}^4) \theta'(u)^2}{n_\mathrm{V}^2 u^6 + (1 + B^2 u^4)  \cos^6\theta(u)}},
\end{equation}
where we substitute the numerical solutions $\theta(u)$ depending on the parameter set ($\pi T/\sqrt{2 B}$, $n_\mathrm{V}/B^{3/2}$).
By numerically evaluating the thermodynamic potential for each solution, the phase diagram for the massless solutions can be constructed, as shown in Figure~\ref{fig:phasediagram} \cite{Evans:2010iy}.

From the perspective of the dual field theory, the Minkowski embedding corresponds to the mesonic phase, where stable meson bound states exist, while the black hole embedding corresponds to the meson-melted phase, where mesonic excitations become dissipative.
The Minkowski embedding supports only vanishing vector charge density, $n_{\rm V}=0$, in the absence of additional charge sources, whereas the black hole embedding allows for a finite $n_{\rm V}$.
In the $\chi$SB phase, one of the low-energy mesonic excitations is identified as a Nambu–Goldstone (NG) mode \cite{Ishigaki:2020vtr}.
As we see later, this NG mode plays a role in the axial current fluctuations associated with the CMW.
Therefore, in this study, we focus on the stable $\chi\text{SB}$ black hole embedding and discuss the CMW by considering fluctuations around this background.


\section{Analysis of QNM of Chiral Cagnetic Wave}\label{sec:4}
In this section, we consider the fluctuations around the background solutions and numerically compute the frequencies of the quasi-normal modes (QNMs). We study the dispersion relation of the CMW and the axial relaxation rate as a function of the background charge density and chiral condensate.
\subsection{Fluctuations}\label{subsec:41}
In this paper, we consider the CMW propagating along the $z$-direction.
Since the CMW arises from the coupling between $U(1)_\mathrm{V}$ and $U(1)_\mathrm{A}$ charges, we introduce the following ansatz for the fluctuations around the background solution:
\begin{equation}\label{eq:ansatz2}
  \begin{gathered}
    \theta= \theta(u) + \delta \theta(t,z,u),\qquad \phi= 0+ \delta \phi (t,z,u) \\
    A_t = a_t(u) + \delta A_t (t,z,u),\qquad A_x =0 ,\qquad A_y = B x,\\
    A_z = 0+ \delta A_z (t,z,u) ,\qquad A_u = 0 + \delta A_u (t,z,u),
  \end{gathered}
\end{equation}
where terms marked with $\delta$ denote fluctuations, and we ignore the fluctuations for $A_{x}$ and $A_{y}$ since they decouple from the other fluctuations in this ansatz.
Expanding the action up to quadratic order in the fluctuations, we obtain the coupled equations of motion for each fluctuation.
We show the explicit forms of the equations in Appendix \ref{sec:A}.
Note that Ref.~\cite{Ishigaki:2020vtr} considered fluctuations depending on the spatial coordinates $(x,y)$ transverse to the magnetic field, in which case $\delta\theta$ and $\delta A_t$ decouple from the other fluctuations.
In contrast, we consider fluctuations propagating along the magnetic field, for which these modes are generally coupled to other fluctuations.

Choosing a gauge of $\delta A_{u} =0$, the equation of motion for $\delta A_{u}$ provides a constraint among $\delta \theta$, $\delta A_t$, and $\delta A_z$.
Performing a Fourier transformation as
\begin{gather}
  \delta \theta (t,z,u)= \wt{\theta}(u) e^{- i (\omega t - k_z z) },\qquad \delta \phi (t,z,u)= \wt{\phi}(u) e^{- i (\omega t - k_z z) } \nn \\
  \delta A_t (t,z,u)= \wt{A}_t(u) e^{- i (\omega t - k_z z) },\qquad \delta A_z (t,z,u)= \wt{A}_z(u) e^{- i (\omega t - k_z z) }, \nn 
\end{gather}
introducing the gauge-invariant combination as
\begin{equation}
\wt{E}_L(u) = \omega\wt{A}_z(u) + k_z \wt{A}_t(u),
\end{equation}
and using the constraint equation, we can rewrite the coupled equations in terms of the three variables $(\wt{\theta},\wt{\phi},\wt{E}_L)$.
For our purpose, we impose the following boundary conditions.
At the black hole horizon, we impose the ingoing-wave boundary condition 
\begin{gather}\label{eq:ingoing}
  \wt{\theta}(u) = f(u)^{-i \frac{\omega u_\mathrm{H}}{4}} \Theta (u),\qquad \wt{\phi}(u) = f(u)^{-i \frac{\omega u_\mathrm{H}}{4}} \Phi (u),\qquad \wt{E}_L(u) = f(u)^{-i \frac{\omega u_\mathrm{H}}{4}} E_L (u),
\end{gather}
where $\Psi\equiv (\Theta,\Phi,{E}_L)^{T}$ is a set of the regular functions at the black hole horizon.

To determine the QNM frequencies, we employ the determinant method discussed in \cite{Kaminski:2009dh}.
We firstly construct three linearly independent solutions $\Psi_{(J)}$ ($J=1,2,3$) satisfying the ingoing-wave boundary condition \eqref{eq:ingoing}. We choose their horizon values as
\begin{equation}
\Psi^{I}{}_{(J)}(u_{\rm H})=\delta^{I}{}_{J},
\qquad I,J\in{1,2,3},
\end{equation}
where $I$ labels the field components and $J$ labels the independent solutions. By numerically integrating the coupled equations of motion from the horizon toward the AdS boundary, we construct the  matrix
\begin{equation}
H^{I}{}_{J}(\omega,k_z,u)=\Psi^{I}{}_{(J)}(u),
\end{equation}
which satisfies $H^{I}{}_{J}(\omega,k_z,u_{\rm H})=\delta^{I}{}_{J}$ by construction.
Defining 
\begin{equation}
    \wh{H}(\omega,k_{z},u) =
    \begin{pmatrix}
        \frac{\cos\theta}{u}& 0 & 0\\
        0 & \frac{\sin\theta}{u}  & 0\\
        0 & 0 & 1
    \end{pmatrix}
    H(\omega,k_{z},u) , 
\end{equation}
the matrix $\wh{H}$ is asymptotically given by
\begin{equation}
\wh{H}^{I}{}_{J}(\omega,k_z,u)
=\mathbb{S}^{I}{}_{J}(\omega,k_z) + {\cal{O}}(u^{2} \log u),
\end{equation}
where $\mathbb{S}$ denotes the coefficient matrix of the non-normalizable modes.
The QNM frequencies are then determined by the vanishing of the determinant of the source matrix,
\begin{equation}
\det\mathbb{S}(\omega,k_z)
=\lim_{u\to0}\det \wh{H}(\omega,k_z,u)=0.
\label{eq}
\end{equation}
By finding the frequencies $\omega$ that satisfy this condition for each momentum $k_z$, we obtain the dispersion relations of the collective excitations.
In the actual calculations, we introduce a small cutoff $u=\epsilon_{\rm UV}$ near the AdS boundary and evaluate the source matrix $\mathbb{S}$ using $\wh{H}(\omega,k_z,\epsilon_{\rm UV})$. We choose $\epsilon_{\rm UV}$ sufficiently small so that the effective quark mass, estimated by $\epsilon_{\rm UV}^{-1}\sin\theta(\epsilon_{\rm UV})$, is negligible.

\subsection{Numerical Analysis}\label{subsec:42}
We now investigate the dispersion relations of the QNMs, focusing on the two lowest modes. Owing to the scaling symmetry of the system, we express all dimensionful quantities in units of the magnetic field hereafter. We also simply denote the momentum $k_{z}$ as $k$ in the following.

\begin{figure}[htbp]
\begin{center}
\includegraphics[width=.9\textwidth]{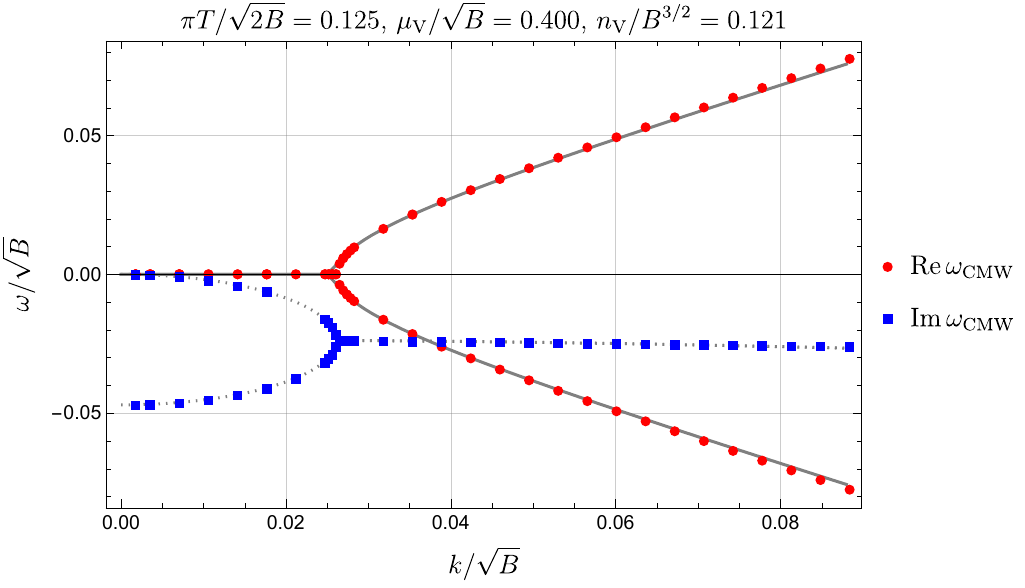}
\caption{Dispersion relations of the QNMs, where filled red circles denote the real part of $\omega$ and filled blue squares denote the imaginary part of $\omega$. The gray curves show fits to (\ref{eq:CMW0}) with $k_c/\sqrt{B} =0.026, \,1/ (\tau \sqrt{B})=0.047$, and $ D\sqrt{B}=0.40$. }
\label{fig:dissOFCMW1}
\end{center}
\end{figure}

Figure~\ref{fig:dissOFCMW1} shows the dispersion relations of the two lowest modes for $n_\mathrm{V}/B^{3/2}=0.121$ and $\pi T /\sqrt{2B}=0.125$.
The filled red circles and blue squares represent the numerical results for the real and imaginary parts of $\omega$, respectively. 
We find that the numerical results are well-fitted by the phenomenological dispersion relation of the CMW (\ref{eq:CMW0}) in the range $0< k/\sqrt{B}<0.1$, with the fitting parameters $k_c/\sqrt{B} =0.026, \,1/ (\tau \sqrt{B})=0.047$, and $ D\sqrt{B}=0.40$.
As discussed in Section \ref{sec:2}, the gapless mode with $\omega_{+}\simeq -i\wt{D}k^{2}$ and the gapped mode with $\omega_{-}\simeq - i/\tau$ emerge in the small-$k$ limit.
For $k<k_c$, both modes have purely imaginary frequencies and exhibit purely damped behavior.
For $k>k_c$, the real parts of their frequencies become finite, indicating a transition to damped oscillatory behavior.

As discussed in \cite{Ishigaki:2020vtr}, the fluctuation $\Phi$ is interpreted as the NG mode arising from the chiral symmetry breaking.
This mode exhibits a diffusive dispersion relation, consistent with the field-theoretical descriptions of dissipative NG modes  \cite{Minami:2015uzo,Hidaka:2019irz,Hongo:2019qhi}.
Furthermore, the NG mode is coupled to the axial current through the Ward identity associated with the axial symmetry \cite{Hoyos:2011us}, and therefore contributes to axial charge fluctuations.
In the presence of a magnetic field, these fluctuations couple to the vector charge sector, described holographically by the longitudinal gauge field fluctuation $\wt{E}_L$.
The resulting coupled collective excitations exhibit a dispersion relation consistent with that of the CMW in the presence of effective relaxation.

\begin{figure}[htbp]
\begin{center}
\includegraphics[width=.95\textwidth]{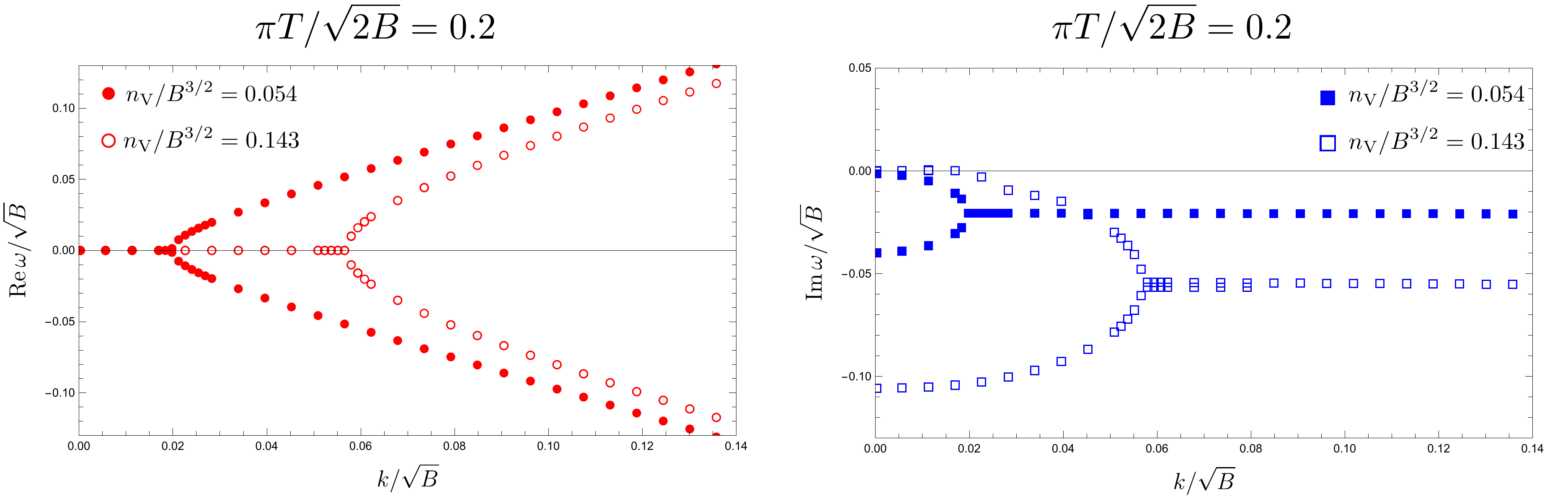}
\caption{
Dependence of the dispersion relation on the charge density $n_\mathrm{V}/B^{3/2}$ at a fixed temperature $\pi T/\sqrt{2B}=0.2$. 
The left panel shows the real part $\mathrm{Re}\,\omega$, and the right panel shows the imaginary part $\mathrm{Im}\,\omega$, where filled symbols correspond to a smaller charge density and open symbols to a larger charge density.
}
\label{fig:ddisscomp}
\end{center}
\end{figure}

We show the dispersion relations for two different values of the charge density $n_\mathrm{V}/B^{3/2}$ with $\pi T/\sqrt{2B}$ fixed in Figure~\ref{fig:ddisscomp}.
The left and right panels represent the real and imaginary part of $\omega$, and filled and open plots represent small and large values of $n_\mathrm{V}/B^{3/2}$, respectively.
The plots reveal that as the charge density increases, the separation between the gapless and gapped modes becomes large, indicating that the relaxation rate $1/\tau$ and the threshold momentum $k_c$ become large.
We also confirm that the numerical results deviate from \eqref{eq:CMW0} as the relaxation rate becomes large.

\begin{figure}[tbp]
\begin{center}
\includegraphics[width=.9\textwidth]{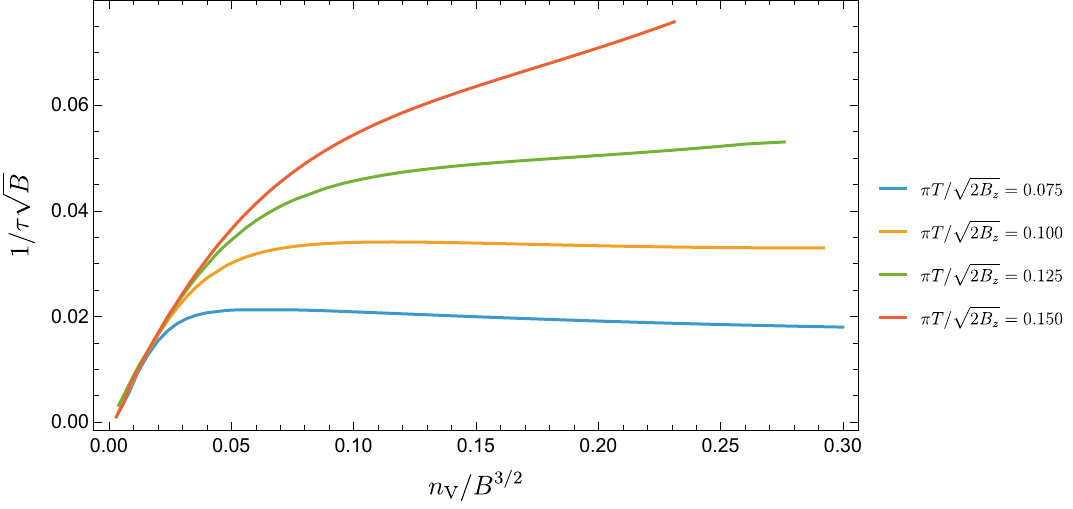}
\caption{
Charge-density dependence of the relaxation rate $1/(\tau \sqrt{B})$ for various fixed temperatures $\pi T/\sqrt{2B}$.}
\label{fig:taunV}
\end{center}
\end{figure}

We also study the relaxation rate $1/(\tau \sqrt{B})$, corresponding to the imaginary part of the gapped mode frequencies in the limit of $k\ra0$, as a function of $n_\mathrm{V}/ B^{3/2}$ in Figure~\ref{fig:taunV}.
Each curve represents the results for different values of $\pi T/\sqrt{2B}$.\footnote{The reason why each plot is cut off partway is that, as can be seen in Figure~\ref{fig:phasediagram}, increasing the charge density any further causes a transition from the $\chi$SB phase to the $\chi$SR phase.}
We find that the relaxation rate approaches zero as the charge density decreases at all temperatures considered. 
When the relaxation rate is sufficiently small, the gapped mode becomes long-lived and can be incorporated into the quasi-hydrodynamic description \cite{Grozdanov:2018fic,Davison:2014lua} together with the gapless mode. In this regime, the dispersion relations of the two modes are well described by the phenomenological expression \eqref{eq:CMW0}.
We also find that in the high-temperature region, the relaxation rate increases as the charge density increases.
On the other hand, in the low-temperature region (specifically at $\pi T/\sqrt{2B} = 0.075$), the relaxation rate attains a maximum at a specific charge density, $n_\mathrm{V}/ B^{3/2}\approx 0.05$, and decreases thereafter.
This demonstrates that while the relaxation time decreases monotonically with charge density at high temperatures, it changes to a non-monotonic behavior at low temperatures.

\begin{figure}[htbp]
\begin{center}
\includegraphics[width=.9 \textwidth]{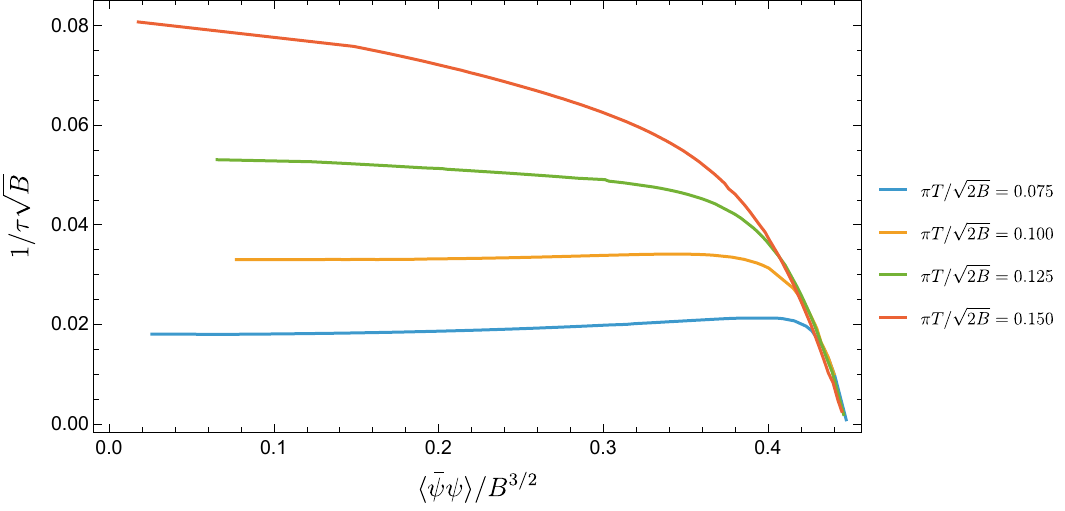}
\caption{
Chiral condensate dependence of the relaxation rate $1/(\tau \sqrt{B})$ for various fixed temperatures $\pi T/\sqrt{2B}$.}
\label{fig:tauCC}
\end{center}
\end{figure}

Furthermore, we investigate the dependence of the relaxation rate on the chiral condensate for several temperatures, as shown in Figure~\ref{fig:tauCC}.
We here emphasize that this dependence is a distinctive feature in our system, where the CMW emerged in the chiral symmetry breaking phase, in contrast to the chirally symmetric phase studied in Ref.~\cite{Kharzeev:2010gd}, where the chiral condensate vanishes.
Similarly to its dependence on the charge density, in the high-temperature regime, the relaxation rate decreases monotonically with increasing chiral condensate, whereas this monotonicity breaks down at lower temperatures.
Across all temperatures investigated, the relaxation rate approaches zero near $\langle \ol{\psi}\psi \rangle/B^{3/2} \approx 0.44$ within our numerical resolution, corresponding to the transition from the dissipative meson-melted phase (black hole embedding) to the dissipationless mesonic phase (Minkowski embedding).

\section{Conclusion and Discussion}\label{sec:5}

In this paper, we investigate the CMW and the associated axial relaxation rate in the D3/D7 model.
In contrast with the original paper \cite{Kharzeev:2010gd}, where the chirally symmetric phase is considered in the Sakai-Sugimoto model, we consider the chiral symmetry breaking phase as a background.
We find that the NG mode associated with the chiral symmetry breaking contributes to the axial current fluctuation.
The resulting collective excitations are consistent with the phenomenological CMW dispersion relation with an effective axial charge relaxation rate, exhibiting a diffusive mode $\omega_{+} \sim - i \wt{D} k^{2}$ and a gapped mode $\omega_{-}\sim -i/\tau$ in the small-$k$ limit.

For small charge density, we find that the dispersion relation is well-matched to the (quasi-)hydrodynamic prediction in \eqref{eq:CMW0}, as shown in figure \ref{fig:dissOFCMW1}, which indicates that the quasi-normal modes we compute are consistent with the chiral magnetic wave in the presence of the axial current relaxation even in the chiral symmetry breaking phase.
Additionally, we reveal the dependence of the relaxation rate ($1/\tau$) as a function of the background charge density and the chiral condensate, corresponding to the strength of the chiral symmetry breaking, for several temperatures as shown in figure \ref{fig:taunV} and figure \ref{fig:tauCC}.
At higher temperatures, the axial relaxation rate decreases monotonically with increasing chiral condensate, whereas this dependence becomes non-monotonic at lower temperatures.
When the axial relaxation rate becomes large, the quasi-hydrodynamic expression is less accurate.
These findings highlight the role of spontaneous chiral symmetry breaking in the low-energy collective dynamics of strongly coupled chiral matter.

Before we conclude this section, we make some remarks on future directions. 
It would be interesting to study the CMW in the chiral symmetry restored phase as well in the D3/D7 model.
Since our results indicate that the phenomenological expression becomes less accurate as the system approaches the symmetry restored phase, the phenomenological description of the CMW as in \eqref{eq:CMW0} may not be directly applicable there in the D3/D7 model in contrast with the Sakai-Sugimoto model.
On the other hand, the emergence of a chiral helix phase has been reported in the chiral symmetry restored phase \cite{Kharzeev:2011rw}, and its possible connection to our findings deserves further investigation.
Another interesting direction is to explore the effect of the explicit symmetry breaking given by the quark mass on the CMW.
Last but not least, applying the external electric field and introducing the axial chemical potential would be an interesting extension.
Since these external fields are known to drive the system to a non-equilibrium steady state in the D3/D7 model \cite{Karch:2007pd,Hoyos:2011us}, it is important to investigate the fate of the CMW in such a steady state.
Furthermore, when the electric field is applied parallel to the magnetic field, the chiral anomaly can induce axial charge production and must therefore be taken into account \cite{Nakamura:2026por}.

\begin{acknowledgments}
The authors would like to thank Shin Nakamura and Shuta Ishigaki for helpful discussions and comments.
The work of MM is supported by Shanghai Sci-tech Co-research Program (Grant No.~25HB2700600) and the Scholarship Fund for Young Researchers from the Promotion and Mutual Aid Corporation for Private Schools of Japan.
\end{acknowledgments}

\appendix
\section{Equations of Motion for the Fluctuations} \label{sec:A}
Expanding the D7-brane action up to quadratic order in fluctuations yields the following expressions:
\begin{align}
\mathcal L^{(2)}_{\rm DBI}
&= -\sqrt{-\det(g+F)}\,\Big[
\tfrac12\Big(\big(1-\theta'^2\gamma^{uu}\big)\gamma^{ab}-\theta'^2S^{au}S^{bu}\Big)\partial_a\delta\theta\,\partial_b\delta\theta
+\tfrac12\sin^2\theta\,\gamma^{ab}\partial_a\delta\phi\,\partial_b\delta\phi \nonumber\\
&\qquad +\gamma^{ac}\gamma^{bd}\,\partial_{[a}\delta A_{b]}\,\partial_{[c}\delta A_{d]}
+\frac{3(3\sin^2\theta-1)}{2\cos^2\theta}\,\delta\theta^2
-3\tan\theta\,\theta'\gamma^{uu}\,\delta\theta\,\partial_u\delta\theta \nonumber\\
&\qquad +3\tan\theta\,S^{ab}\partial_a\delta A_b\,\delta\theta
+2\theta'\gamma^{ac}S^{ub}\,\partial_c\delta\theta\,\partial_{[a}\delta A_{b]}\Big],\\[4pt]
\mathcal L^{(2)}_{\rm WZ}
&= 2B\cos^4\theta\left(\partial_t\delta\phi\,\partial_{[z}\delta A_{u]}+\partial_z\delta\phi\,\partial_{[u}\delta A_{t]}+\partial_u\delta\phi\,\partial_{[t}\delta A_{z]}\right) \nonumber\\
&\quad -4Ba_t'\cos^3\theta\sin\theta\,\delta\theta\,\partial_z\delta\phi .
\end{align}
where, $\gamma^{ab}$ $(a,b=t,z,u)$ and $S^{ab}$ denote the symmetric and antisymmetric components of $(g+F)^{-1}$.
$\gamma _{ab}$ is the open string metric, defined by $\gamma_{ab}= g_{ab} -(F g^{-1}F)_{ab}$ \cite{Kim:2011qh,Sonner:2012if,Nakamura:2013yqa}.
Varying this Lagrangian yields the following equations of motion for the fluctuations:
\begin{align}
&\partial_a\!\left(\sqrt{-\det(g+F)}\,\sin^2\theta\,\gamma^{ab}\partial_b\delta\phi\right)
+4B\cos^3\theta\sin\theta\left(2\theta'\,\partial_{[t}\delta A_{z]}+a_t'\,\partial_z\delta\theta\right)=0,\\[6pt]
&\partial_a\!\left[\sqrt{-\det(g+F)}\left(\Big(\big(1-\theta'^2\gamma^{uu}\big)\gamma^{ab}-\theta'^2S^{au}S^{bu}\Big)\partial_b\delta\theta
+2\theta'\gamma^{ab}S^{uc}\,\partial_{[b}\delta A_{c]}\right)\right]\nonumber\\
&\qquad+\left[\frac{3(1-3\sin^2\theta)}{\cos^2\theta}\sqrt{-\det(g+F)}
-\partial_u\!\left(3\sqrt{-\det(g+F)}\,\tan\theta\,\theta'\gamma^{uu}\right)\right]\delta\theta\nonumber\\
&\qquad-3\sqrt{-\det(g+F)}\,\tan\theta\,S^{ab}\partial_a\delta A_b
-4Ba_t'\cos^3\theta\sin\theta\,\partial_z\delta\phi=0,\\[6pt]
&2\partial_a\!\left[\sqrt{-\det(g+F)}\left(\gamma^{ac}\gamma^{bd}\,\partial_{[c}\delta A_{d]}
-\theta'\gamma^{c[a}S^{b]u}\,\partial_c\delta\theta
+\tfrac32\tan\theta\,S^{ab}\delta\theta\right)\right]\nonumber\\
&\qquad+8B\theta'\cos^3\theta\sin\theta\,\delta^{[b}_t\delta^{c]}_z\,\partial_c\delta\phi=0 .
\end{align}



 \bibliographystyle{JHEP}
 \bibliography{CMW}

\end{document}